\documentclass[11pt]{article}

\usepackage{amssymb} 
\usepackage{amsmath} 
\usepackage{amsthm} 
\usepackage{setspace} 
\usepackage{cite} 
\usepackage{hyperref}
\usepackage{tikz}
\usepackage{verbatim}
\usepackage{graphicx}
\usepackage{subfig}
\usetikzlibrary{shapes.geometric}

\newtheorem{definition}{Definition} 
\newtheorem{corollary}{Corollary}

\newtheorem{lemma}{Lemma} 
\newtheorem{theorem}{Theorem} 
 
\newtheorem{obs}{Observation}

\newtheorem{remark}{Remark}

\newcommand{\F}{\mathbb{F}}

\newcommand{\C}{\mathbb{C}}

\newcommand{\Q}{\mathbb{Q}}
\newcommand{\R}{\mathbb{R}}

\newcommand{\depththree}{\Sigma\Pi\Sigma} 
\newcommand{\depthfour}{\Sigma\Pi\Sigma\Pi}

\newcommand{\pit}{\mbox{\small\rm PIT}} 
\newcommand{\IP}{\mbox{\small\rm IP}} 
\newcommand{\PSPACE}{\mbox{\small\rm PSPACE}} 
\newcommand{\corp}{\mbox{\small\rm co-RP}} 
\renewcommand{\char}{\mbox{\small\rm Char}}

\DeclareMathOperator{\poly}{\mbox{\small\rm poly}} 
  
  \DeclareMathOperator{\perm}{\mbox{\small\rm Perm}}

\title{A Note on Polynomial Identity Testing for Depth-3 Circuits}  

\author{V. Arvind\thanks{Institute of Mathematical Sciences, Chennai,
    India, \texttt{email: arvind@imsc.res.in}}  \and Abhranil Chatterjee \thanks{Institute of Mathematical Sciences, Chennai,
    India, \texttt{email: abhranilc@imsc.res.in}} \and Rajit Datta\thanks{Chennai Mathematical Institute, Chennai, India, \texttt{email: rajit@cmi.ac.in}} \and Partha
  Mukhopadhyay\thanks{Chennai Mathematical Institute, Chennai, India,
    \texttt{email: partham@cmi.ac.in}}
}

\begin{document}

\maketitle

\begin{abstract}
Let $C$ be a depth-3 arithmetic circuit of size at most $s$, computing a polynomial $f\in\F[x_1,\ldots, x_n]$ (where $\F$ = $\Q$ or $\C$) 
and the fan-in of the product gates of $C$ is bounded by $d$. We give a deterministic polynomial identity testing algorithm to check whether $f\equiv 0$ or not in time $2^d \poly(n,s)$.  

Over finite fields, for $\char(\F) > d$ we give a deterministic algorithm of running time $2^{\gamma\cdot d}\poly(n,s)$ where $\gamma\leq 5$. 
\end{abstract}

\section{Introduction}
Polynomial Identity Testing ($\pit$) is the following problem :  Given an arithmetic circuit $C$ computing a polynomial in $\F[x_1, \ldots, x_n]$, determine whether $C$ computes an identically zero polynomial or not.  
The problem can be presented either in the \emph{white-box} model or in the \emph{black-box} model.  In the white-box model, the arithmetic circuit is given explicitly as the input. In the black-box model, the arithmetic circuit is given black-box access, and the circuit can be evaluated over any point in $\F^n$ (or $F^n$ where $\F \subseteq F$ is an extension field).
Over the years, the problem has played pivotal role in many important results in complexity theory and algorithms: Primality Testing \cite{aks04},  the PCP Theorem \cite{almss98}, $\IP=\PSPACE$ \cite{shamir90}, graph matching algorithms 
\cite{lov79,mvv87}. The problem $\pit$ admits a $\corp$ algorithm via the Schwartz-Zippel-Lipton-DeMillo Lemma 
\cite{Sch80,Zip79,DL78}, but an efficient deterministic algorithm is known only in some special cases. 
An important result of Impagliazzo and Kabanets \cite{KI04} (also, see \cite{HS80, Agr05}) shows a connection between the existence of a subexponential time $\pit$ algorithm and arithmetic circuit lower bounds. 
We refer the reader to the survey of Shpilka and Yehudayoff \cite{sy10} for the exposition of important results in arithmetic circuit complexity, and polynomial identity testing problem. 

In a surprising result, Agrawal and Vinay \cite{av2008} show that an efficient deterministic $\pit$ algorithm only for depth-4 $\depthfour$ circuits is sufficient for obtaining an efficient deterministic $\pit$ algorithm for the general 
arithmetic circuits. The main technical ingredient in their proof is an ingenious depth-reduction technique. 
Over characteristic zero fields, derandomization of $\pit$ even for depth-3 $\Sigma\Pi\Sigma$ circuits 
suffices \cite{gkks2013}. 

Motivated by the results of \cite{KI04,Agr05,av2008}, a large body of research consider the polynomial identity testing problem for restricted classes of depth-3 and depth-4 circuits. A particularly popular model in depth three arithmetic circuits is $\depththree(k)$ circuit, where the fan-in of the top $\Sigma$ gate is bounded by $k$. Dvir and Shpilka have shown a \emph{white-box} quasi-polynomial time deterministic $\pit$ algorithm for $\depththree(k)$ circuits \cite{DS07}. Kayal and Saxena have given a deterministic $\poly(d^k,n,s)$ white-box 
algorithm for the same problem \cite{KS07}. Following the result of \cite{KS07}(Also see \cite{AM10} for a different analysis). 
Karnin and Shpilka have given the first \emph{black-box} quasi-polynomial time algorithm for $\depththree(k)$ circuits \cite{KS11}. Later, Kayal and Saraf  \cite{KS09} have shown a polynomial-time deterministic black-box $\pit$ algorithm for the same class of circuits over $\Q$ or $\R$. Finally, Saxena and Sheshadhri have settled the situation completely by giving a deterministic polynomial-time \emph{black-box} algorithm for $\depththree(k)$ circuits \cite{SS12} over any field. Recently, Oliveira et al. have given a sub-exponential $\pit$-algorithm for depth-3 and depth-4 \emph{multilinear} formulas \cite{OSV16}.

For general depth-3 $\Sigma\Pi\Sigma$ circuits with $\times$-gate fan-in bounded by $d$ no deterministic algorithm with running time better than $\min\{ d^n, n^d \} \poly(n,d)$ is known. Our main results are the following. 

\begin{theorem}\label{main-thm}
Let $C$ be a depth-3 $\Sigma\Pi\Sigma$ circuit of size at most $s$, computing a polynomial $f \in \F[x_1,\ldots, x_n]$ ( where $\F$ = $\Q$ or $\C$) and the fan-in of the product gates of $C$ is bounded by $d$. We give a \emph{white-box} deterministic polynomial time identity testing algorithm to check whether $f\equiv 0$ or not in time $2^d \poly(n,s)$.  
\end{theorem}

As an immediate corollary we get the following. 

\begin{corollary}\label{poly-depth3}
Let $C$ be a depth-3 $\Sigma\Pi\Sigma$ circuit of size at most $s$, computing a polynomial $f\in \F[x_1,\ldots, x_n]$ ( where $\F$ = $\Q$ or $\C$) and the fan-in of the product gates of $C$ is bounded by $O(\log n)$. We give a deterministic $\poly(n,s)$ time identity testing algorithm to check whether $f\equiv 0$ or not.  
\end{corollary}

Over the fields of positive characteristics, we show the following result. 

\begin{theorem}\label{main-thm-2}
Let $C$ be a depth-3 $\Sigma\Pi\Sigma$ circuit of size at most $s$, computing a polynomial $f \in \F[x_1,\ldots, x_n]$ and the fan-in of the product gates of $C$ is bounded by $d$. For $\char(\F) > d$, we give a \emph{white-box} deterministic polynomial time identity testing algorithm to check whether $f\equiv 0$ or not in time $2^{\gamma\cdot d} \poly(n,s)$. 
The constant $\gamma$ is at most $5$. 
\end{theorem}

\section{Orgazination}
The paper is organized as follows. Section \ref{prelim} contains preliminary materials. In Section \ref{main-sec}, we prove Theorem \ref{main-thm}. Theorem \ref{main-thm-2} is proved in Section \ref{main-sec-1}. 


\section{Preliminaries}\label{prelim}
For a monomial $m$ and a polynomial $f$, let $[m]f$ denote the coefficient of the monomial $m$ in $f$.
We denote the field of rational numbers as $\mathbb{Q}$, and the field of complex numbers as $\mathbb{C}$. The depth-3 $\Sigma\Pi\Sigma(s,d)$ circuits compute polynomials of the following form:
\[
C(x_1,\ldots,x_n) = \sum_{i=1}^s \prod_{j=1}^d L_{i,j}(x_1,\ldots,x_n).
\]
where $L_{i,j}$'s are affine linear forms over $\F$. The following observation is well-known and it says that for $\pit$ purpose it is sufficient to consider homogeneous circuits. 

\begin{obs}
Let $C(x_1,\ldots,x_n)$ be a $\Sigma\Pi\Sigma(s,d)$ circuit. Then $C\equiv 0$ if and only if 
$z^d C(x_1/z, \ldots,x_n/z)\equiv 0$ where $z$ is a new variable. 
\end{obs}
 
We use the notation $\Sigma^{[s]} \Pi^{[d]} \Sigma$ to denote homogeneous depth-3 circuits of top $\Sigma$ gate fan-in $s$, product gates fan-in bounded by $d$. 


We recall the definition of Hadamard Product of two polynomials. The concept of Hadamard product  is particularly useful in noncommutative computations \cite{AJ09,AS18}. 
\begin{definition}\label{hadprod}
 Given two degree $d$ polynomials $f,g \in \F[x_1,x_2,\ldots,x_n]$,  the Hadamard Product $f \circ g$ is defined as
 
 $$f \circ g = \sum_{m : \deg(m)\leq d} ([m]f \cdot [m]g) \ m.$$
 
\end{definition}

For the $\pit$ purpose in the commutative setting, we adapt the notion of Hadamard Product suitably and define a scaled version of Hadamard Product of two polynomials. 
\begin{definition}\label{s-hadprod}
 Given two degree $d$ polynomials $f,g \in \F[x_1,x_2,\ldots,x_n]$,  the scaled version of the Hadamard Product  $f \circ^{s} g$ is defined as
 
 $$f \circ^{s} g = \sum_{m : \deg(m)\leq d} (m! \cdot [m]f \cdot [m]g) \ m$$
 
\end{definition}

where $m=x^{e_1}_{i_1}x^{e_2}_{i_2} \ldots x^{e_r}_{i_r}$ for some $r\leq d$ and by abusing the notation we define $m! = e_1 ! \cdot e_2! \cdot \ldots \cdot e_r!$.

For the purpose of $\pit$ over $\Q$, it is enough to be able to compute $f \circ^{s} f (1,1,\ldots,1)$. 
As $f \circ^{s} f$ has only non-negative coefficients, we will see a non-zero value when we compute $f \circ^{s} f (1,1,\ldots,1)$ if and only if $f \not \equiv 0$.
Over $\C$ it is enough to compute $f \circ^{s} \bar{f} (1,1,\ldots,1)$ where $\bar{f}$ denotes the polynomial obtained by conjugating every coefficient of $f$.

We also recall a result of Ryser~\cite{ry63} that gives a $\Sigma^{[2^n]} \Pi^{[n]} \Sigma$ circuit for the Permanent polynomial of $n\times n$ symbolic matrix.

\begin{lemma}[Ryser ~\cite{ry63}]\label{ryser}
For a matrix $X$ with variables $x_{ij} : 1\leq i,j\leq n$ as entries,
$$\perm(X)= (-1)^n \sum_{S \subseteq [n]} (-1)^{|S|} \prod^{n}_{i=1} \left(\sum_{j \in S} x_{ij}\right).$$
\end{lemma}

\begin{lemma}\label{calc}
 For a monomial $m = x_{i_1}x_{i_2} \ldots x_{i_d}$ ($i_1 ,\ldots , i_d$ need not be distinct) and a homogeneous $\Pi\Sigma$ circuit $C=\prod^d_{j=1} L_j$ we have:
 
 \[
  [m]C =  \frac{1}{m!} \sum_{\sigma \in S_d} \prod^d_{j=1}( [x_{i_j}] L_{\sigma(j)}).
 \]
\end{lemma}
\begin{proof}

The monomial $m$ can be obtained from $C$ by first fixing a bijection $\sigma : [d] \mapsto [d]$ and considering the coefficient $[m]C_{\sigma}= \prod^{d}_{j=1} [x_{i_{\sigma(j)}}] L_j = \prod^{d}_{j=1} [x_{i_{j}}] L_{\sigma^{-1}(j)}$.
This is one way of generating this monomial and this monomial $m$ can be generated in many different orders. The final $[m]C$ is the sum of all coefficients $[m]C_{\sigma}$ generated in all distinct orders.

Now if $m=x^{e_1}_{i_1}x^{e_2}_{i_2} \ldots x^{e_r}_{i_r}$ for some $r\leq d$ then for a fixed $\sigma$  one can obtain $m!$ different bijections that do not change the string $x_{i_{\sigma(1)}}x_{i_{\sigma(2)}} \ldots x_{i_{\sigma(d)}}$ and these will generate
the same coefficient $\prod^{d}_{j=1} [x_{i_{\sigma(j)}}] L_j$ . Thus only the bijections that produce a different string from $x_{i_{\sigma(1)}}x_{i_{\sigma(2)}} \ldots x_{i_{\sigma(d)}}$ are relevant
($[m]C_{\sigma} = [m]C_{\pi}$ if the strings $x_{i_{\sigma(1)}}x_{i_{\sigma(2)}} \ldots x_{i_{\sigma(d)}}$ and $x_{i_{\pi(1)}}x_{i_{\pi(2)}} \ldots x_{i_{\pi(d)}}$ are identical). To account for the coefficients produced by
the extra bijections we divide by $m!$

 

 

\end{proof}

Now we are ready to prove the main theorems.

\section{The results over zero characteristics}\label{main-sec}

To prove Theorem \ref{main-thm}, the following theorem is sufficient. 

\begin{theorem}\label{main}
 Given a homogeneous $\Sigma^{[s]} \Pi^{[d]} \Sigma$ circuit $C$ computing a degree $d$ polynomial in  $\F[x_1 ,x_2 ,\ldots , x_n]$
 (where $\F$ = $\Q$ or $\C$), we can test whether $C \equiv 0$ or not deterministically in $2^{d} \poly(s,n)$ time.
\end{theorem}

\begin{proof}
 For simplicity, we present the proof only over $\Q$. Over $\C$, we need a minor modification as explained in Remark \ref{rmk1}. 
 Given the circuit $C$ we compute $C\circ^{s} C$ and evaluate at $(1,1,\ldots, 1)$ point. Notice that over rationals, $C \circ^{s} C$ has non-negative coefficients. This also implies that $C\equiv 0$ 
 if and only if $C\circ^{s} C(1,1,\ldots,1) = 0$. So it is sufficient to show that $C\circ^{s} C(1,\ldots,1)$ can be computed deterministically in time $2^d \poly(s,n)$. 
 Since the scaled Hadamard Product distributes over addition, we only need to show that the scaled Hadamard Product of 
 two $\Pi\Sigma$ circuits can be computed efficiently.
 
 \begin{lemma}\label{thetrick}
  Given two homogeneous $\Pi^{[d]} \Sigma$ circuits $C_1 =
  \prod^d_{i=1} L_i$ and $C_2 = \prod^d_{i=1} {L'}_i$ we have:
 \[
C_1 \circ^{s} C_2 = \sum_{\sigma \in S_d} \prod^d_{i=1} (L_i \circ^{s} {L'}_{\sigma(i)}).
\]
 \end{lemma}

 \begin{proof}
 We prove the formula monomial by monomial. Let $m= x_{i_1} x_{i_2}
 \ldots x_{i_d}$ be a monomial in $C_1$ (Note that $i_1 ,
 i_2,\ldots,i_d$ need not be distinct).
 
 Now let $m$ be a monomial that appears in both $C_1$ and $C_2$. From Lemma~\ref{calc} the coefficients are $$[m]C_1 = \alpha_1 = \frac{1}{m!} \left(\sum_{\sigma \in S_d} \prod^d_{j=1}
 [x_{i_j}] {L}_{\sigma(j)}\right)$$ and $$[m]C_2 = \alpha_2 = \frac{1}{m!} \left(\sum_{\pi \in S_d} \prod^d_{j=1}
 [x_{i_j}] {L'}_{\pi(j)}\right)$$ respectively.

 From the definition~\ref{s-hadprod} we have $$[m] (C_1 \circ^s C_2)=m! \cdot \alpha_1 \cdot \alpha_2.$$
 
Now let us consider the matrix $T$ where $T_{ij} = L_i \circ^s {L'}_j :
1\leq i,j\leq d$ and $\perm(T)= \sum_{\sigma \in S_d} \prod^d_{i=1}
L_i \circ^s {L'}_{\sigma(i)}$.  The coefficient of $m$ in $\perm(T)$ is
 
 $$ [m]\perm(T) = \sum_{\sigma \in S_d} [m] \left(\prod^d_{j=1} L_j \circ^s
{L'}_{\sigma(j)}\right).$$ 

Similar to Lemma~\ref{calc}, we notice the following.
 
 $$[m]\perm(T)=\sum_{\sigma \in S_d} \frac{1}{m!} \sum_{\pi \in S_d} \prod^d_{j=1} [x_{i_j}] (L_{\pi(j)} \circ^s {L'}_{\sigma(\pi(j))})$$
 $$= \frac{1}{m!} \sum_{\sigma \in S_d} \sum_{\pi \in S_d} \prod^d_{j=1} ([x_{i_j}] L_{\pi(j)}) \cdot ([x_{i_j}] {L'}_{\sigma(\pi(j))})$$
 $$=\frac{1}{m!}\sum_{\sigma \in S_d} \sum_{\pi \in S_d} \prod^d_{j=1} ([x_{i_j}] L_{\pi(j)}) \cdot \prod^d_{j=1} ([x_{i_j}] {L'}_{\sigma(\pi(j))})$$
 $$=\sum_{\pi \in S_d} \left(\prod^d_{j=1} ([x_{i_j}] L_{\pi(j)}) \cdot \frac{1}{m!} \sum_{\sigma \in S_d} \prod^d_{j=1} ([x_{i_j}] {L'}_{\sigma(\pi(j))})\right)$$
 $$= m! \cdot \frac{1}{m!}  \sum_{\pi \in S_d} \left(\prod^d_{j=1} ([x_{i_j}] L_{\pi(j)}) \cdot \frac{1}{m!} \sum_{\sigma \in S_d} \prod^d_{j=1}( [x_{i_j}] {L'}_{\sigma(\pi(j))})\right).$$
 
Clearly, for any fixed $\pi\in S_d$, we have that $\sum_{\sigma \in
  S_d} \prod^d_{j=1} [x_{i_j}] {L'}_{\sigma(\pi(j))}= m! \alpha_2$. Hence,
$[m]\perm(T) = m! \cdot \alpha_1 \cdot \alpha_2$ and the lemma follows.

 \end{proof}

\begin{lemma}\label{d2}
 Given two $\Pi^{[d]} \Sigma$ circuits $C_1$ and $C_2$ we can compute
 a $\Sigma^{[2^d]} \Pi^{[d]} \Sigma$ for $C_1 \circ^s C_2$ in time $2^d
 \poly(n,d)$.
\end{lemma}

\begin{proof}
From Lemma~\ref{thetrick} we observe that $\perm(T)$ gives a circuit
for $C_1 \circ^s C_2$.  A $\Sigma^{[2^d]} \Pi^{[d]} \Sigma$ circuit for
$\perm(T)$ can be computed in $2^d \poly(n,d)$ time using
Lemma~\ref{ryser}.

\end{proof}

Now we show how to take the scaled Hadamard Product of two $\Sigma \Pi \Sigma$ circuits.

\begin{lemma}\label{d3}
 Given two $\Sigma \Pi^{[d]} \Sigma$ circuits $C = \sum^{s}_{i=1} P_i$
 and $\widetilde{C} = \sum^{\tilde{s}}_{i=1} \widetilde{P_i}$ We can
 compute a $\Sigma^{[2^d s \tilde{s}]} \Pi^{[d]} \Sigma$ circuit for
 $C \circ^s \widetilde{C}$ in time $2^d \poly(s,\tilde{s},d,n)$.
\end{lemma}

\begin{proof}
We first note that by distributivity,
$$C \circ^s \widetilde{C} = \sum^{s}_{i=1} \sum^{\tilde{s}}_{j=1} P_i
\circ^s \widetilde{P_j}.$$

Using Lemma~\ref{d2} for each pair $P_i \circ^s \widetilde{P_j}$ we get
a $\Sigma^{[2^d]} \Pi^{[d]} \Sigma$ circuit $P_{ij}$.  Now the formula
$\sum^{s}_{i=1} \sum^{\tilde{s}}_{j=1} P_{ij}$ is a $\Sigma^{[2^d s
    \tilde{s}]} \Pi^{[d]} \Sigma$ formula which can be computed in
$2^d \poly(s,\tilde{s},d,n)$ time.

\end{proof}

Now given a $\Sigma^{[s]} \Pi^{[d]} \Sigma$ circuit $C$ we can compute
$C \circ^s C$ using Lemma~\ref{d3} and finally evaluating $C \circ^s C
(1,1,\ldots,1)$ completes the $\pit$ algorithm. Clearly all the
computation can be done in $2^{d} \poly(s,n)$ time. This completes the
proof of Theorem~\ref{main}.

\end{proof}

\begin{remark}\label{rmk1}
 To adapt the algorithm over $\mathbb{C}$, we need to just compute $C
 \circ^s \bar{C}$ where $\bar{C}$ is the polynomial obtained from $C$ by
 conjugating each coefficient. Note that a circuit computing $\bar{C}$
 can be obtained from $C$ by just conjugating the scalars that appear
 in the linear forms of $C$. This follows from the fact that the
 conjugation operation distributes over addition and multiplication.
 Now we have $[m] (C \circ^s \bar{C}) ={|[m]C|}^2$, so the coefficients
 are all positive and thus evaluating $C \circ^s \bar{C}(1,1,\ldots,1)$
 is sufficient for the PIT algorithm.
\end{remark}

\section{The results over finite fields}\label{main-sec-1}

In this section we extend the $\pit$ results over the finite fields. 
Now we state the main theorem of the section. 

\begin{theorem}\label{main-thm-2}
Let $C$ be a depth-3 $\Sigma\Pi\Sigma$ circuit of size at most $s$, computing a polynomial $f \in \F[x_1,\ldots, x_n]$ and the fan-in of the product gates of $C$ is bounded by $d$. For $\char(\F) > d$, we give a \emph{white-box} deterministic polynomial time identity testing algorithm to check whether $f\equiv 0$ or not in time $2^{\gamma\cdot d} \poly(n,s)$. 
The constant $\gamma$ is at most $5$. 
\end{theorem}

\begin{proof}
Consider first the case when $p=\char(\F) > d$.  
From Lemma \ref{calc}, notice that for any $\Pi^{[d]}\Sigma$ circuit $P$, 
\[
[m]P =  \frac{1}{m!} \sum_{\sigma \in S_d} \prod^d_{j=1}( [x_{i_j}] L_{\sigma(j)}).
\]
and $m!\neq 0 \mod p$. Now define the $d\times d$ matrix $T_P$ such that each row of $T_P$ is just the linear forms 
$L_1 L_2 \ldots L_d$ appearing in $P$ \footnote{Again, we identify the linear forms as $L_{1}, L_2, \ldots, L_d$ where $L_1, \ldots,L_{e_1}$ are the same, $L_{e_1 + 1}, \ldots, L_{e_1 + e_2}$ are the same and so on.}. Clearly the following is true. 
\[
\perm(T_P)=\sum_{\sigma \in S_d} \prod^d_{j=1} L_{\sigma(j)}.
\]
Use Ryser's formula given by Lemma \ref{ryser}, to express $\perm(T_P)$ as a depth-3 $\Sigma^{[2^d]}\Pi^{[d]}\Sigma$ circuit. If $C=P_1 + \ldots + P_s$, consider the polynomial $f_C = \sum_{i=1}^s \perm(T_{P_i})$. Notice that $f_C$ can be expressed as $\Sigma^{[2^d \cdot s]}\Pi^{[d]}\Sigma$ circuit. Consider the \emph{noncommutative} version of the polynomial $f_C$ which we denote as $f_C^{nc}$. Clearly we have a noncommutative ABP for $f_C^{nc}$ of width $w=2^d \cdot s$ and $d$ many layers.

%
Now we make an important observation from the proof of Lemma \ref{calc}. Suppose $\mathcal{M}$ be the set of all monomials of degree $d$ over $x_1,\ldots,x_n$. For a fixed monomial $m\in\mathcal{M}$ of form $x_{i_1}x_{i_2}\ldots x_{i_d}$ where $i_1\leq i_2\leq\ldots \leq i_d$ and $\sigma\in S_d$, define $m^\sigma = x_{i_{\sigma(1)}}x_{i_{\sigma(2)}}\ldots x_{i_{\sigma(d)}}$. The monomial $m$ can be present in $f_C^{nc}$ in different orders $m^\sigma$. We claim that $f\equiv 0$ if and only if $f_C^{nc}\equiv 0$. To see the claim, the following simple lemma suffices.

\begin{lemma}\label{important}
Let $f = \sum_{m\in\mathcal{M}} [m]f \cdot {m}$ where $[m]f \in\F$ for all monomials $m\in\mathcal{M}$. Then 
\[f_C^{nc} = \sum_{m\in\mathcal{M}} \sum_{\sigma\in S_d} m! \cdot [m]f \cdot m^{\sigma}.\] 
\end{lemma}

\begin{proof}
Let $x_{i_1}\ldots x_{i_d}$ be a fixed ordering of a monomial $m$ appearing in $f_{C}^{nc}$. The coefficient of $x_{i_1}\ldots x_{i_d}$ in $\perm(T_P)=\sum_{\sigma \in S_d} \prod^d_{j=1} L_{\sigma(j)}$ is simply 
$\sum_{\sigma\in S_d} \prod_{j=1}^d [x_{i_j}]L_{\sigma(j)}.$ 
But from Lemma \ref{calc},
$\sum_{\sigma\in S_d} \prod_{j=1}^d [x_{i_j}]L_{\sigma(j)}$  is exactly 
$m! \cdot [m]P$.
 Since $[m]f = \sum_{i=1}^s [m]{P_i}$, the lemma follows. 
  
\end{proof}

Now we apply the identity testing algorithm of 
Raz and Shpilka for noncommutative ABPs on the ABP of $f_C^{nc}$ to get the desired result \cite{RS05}. The bound on $\gamma$ comes from Theorem 4 of their paper \cite{RS05}.  

%

 


\end{proof}

As an immediate application of Theorem \ref{main-thm-2}, we state the following corollary. 

\begin{corollary}\label{poly-depth3-finite}
Let $C$ be a depth-3 $\Sigma\Pi\Sigma$ circuit of size at most $s$, computing a polynomial $f\in \F[x_1,\ldots, x_n]$ and the fan-in of the product gates of $C$ is bounded by $d$. Suppose 
that $\char(\F) > d$. For $d=O(\log n)$, we give a deterministic $\poly(n,s)$ time identity testing algorithm to check whether $f\equiv 0$ or not. 
\end{corollary}


\newcommand{\etalchar}[1]{$^{#1}$}

\end{document}